\documentclass[aps,prx,reprint,superscriptaddress]{revtex4-2}
\usepackage[T1]{fontenc}
\usepackage{hyperref}
\hypersetup{
  colorlinks   = true,
  urlcolor     = blue,
  linkcolor    = blue,    
  citecolor    = blue
}

\usepackage{siunitx}
\usepackage{physics,amsfonts,amsmath,graphicx,dcolumn,bm,array,color,amssymb,setspace,empheq,braket,multirow,cases,enumitem}
\usepackage[ruled,lined]{algorithm2e}

\begin{document}

\title{Device for MHz-rate rastering of arbitrary 2D optical potentials}

\author{Edita Bytyqi}
\affiliation{
 Department of Physics, MIT-Harvard Center for Ultracold Atoms and Research Laboratory of Electronics, Massachusetts Institute of Technology, Cambridge, Massachusetts 02139, USA\\
 }
\author{Josiah Sinclair}
\affiliation{
 Department of Physics, MIT-Harvard Center for Ultracold Atoms and Research Laboratory of Electronics, Massachusetts Institute of Technology, Cambridge, Massachusetts 02139, USA\\
 }
 \affiliation{
 Present address: Department of Physics, University of Wisconsin-Madison, 1150 University Avenue, Madison, WI, 53706, USA\\
 }
\author{Joshua Ramette}
\affiliation{
 Department of Physics, MIT-Harvard Center for Ultracold Atoms and Research Laboratory of Electronics, Massachusetts Institute of Technology, Cambridge, Massachusetts 02139, USA\\
 }
\author{Vladan Vuleti\'c}
\email{vuletic@mit.edu}
\affiliation{
 Department of Physics, MIT-Harvard Center for Ultracold Atoms and Research Laboratory of Electronics, Massachusetts Institute of Technology, Cambridge, Massachusetts 02139, USA\\
 }
\date{\today}

\begin{abstract}
Current architectures for neutral-atom arrays utilize devices such as acousto-optic deflectors (AODs) and spatial light modulators (SLMs) to multiplex a single classical control line into $N$ qubit control lines. Dynamic control is speed-limited by the response time of AODs, and geometrically constrained to respect a product structure, limiting motion to row-by-row or column-by-column moves. We propose an optical rastering device that can produce any 2D pattern, not limited to grids, at 1 MHz refresh rates. We demonstrate a design with a resolution of 40 x 40 that can be further scaled up to 100 x 100 to match existing and future neutral atom devices. The ability to simultaneously transport atomic qubits in arbitrary directions will enhance qubit connectivity, enable more efficient circuits, and may have broader applications ranging from LiDAR to fluorescence microscopy.

\end{abstract}

\maketitle
Optical beam deflectors are foundational components in a wide range of applications spanning physics, biology, and emerging optical technologies. In two-photon fluorescence microscopy, state-of-the-art two-dimensional (2D) deflectors enable rapid volumetric imaging of neural activity by steering tightly focused laser beams at high speeds \cite{reddy_three-dimensional_2008, wu_kilohertz_2020, wu_multi-mhz_2017, wu_ultrafast_2017, lai_high-speed_2021}. Such devices have also been critical to the development of LiDAR systems for autonomous vehicles and other advanced sensing applications \cite{li_virtually_2021}. More recently, 2D optical deflectors have become essential tools in quantum computing, where they are used to generate and manipulate arrays of individually trapped neutral atoms, enabling a flexible and scalable platform for quantum information processing \cite{bluvstein_logical_2024, evered_high-fidelity_2023, manetsch_tweezer_2024}.

A key advantage of neutral-atom arrays as a quantum computing platform is their scalability and high-degree of connectivity, achieved by transporting optically trapped atoms across a plane by means of 2D acousto-optic deflectors (2D AODs). However, in those implementations the two axes are coupled, restricting the parallel shuttling of atoms to transport along individual rows or columns. This geometric restriction limits the degree of parallelism in rearrangement protocols, ultimately increasing the time overhead required for reconfiguration in large-scale systems.

In the present work, we introduce a novel Megahertz-rate rastering device for pulsed generation of arbitrary 2D optical potentials. 
Employing a dual-axis approach — with a "fast" axis used for tone switching and a "slow" axis for scanning — we demonstrate arbitrary 2D intensity patterns at 1 MHz refresh rates with a $40 \times 40$ resolution. 
The scan time along the slow axis is 1~$\mu$s and the switching time along the fast axis is below 5~ns. 
The raster rate $R_{\text{raster}}$ is designed to be much larger than twice the typical atom trapping frequency of $f_{\text{trap}}=50$~kHz, $R_{\text{raster}}>>2\cdot f_{\text{trap}}$, to avoid parametric heating \cite{hutzler_2017, savard_heating_1997, zheng_heating_2022}.

\begin{figure*}
    \centering
    \includegraphics[width=\linewidth,trim=0cm .25cm 0cm .7cm,clip]{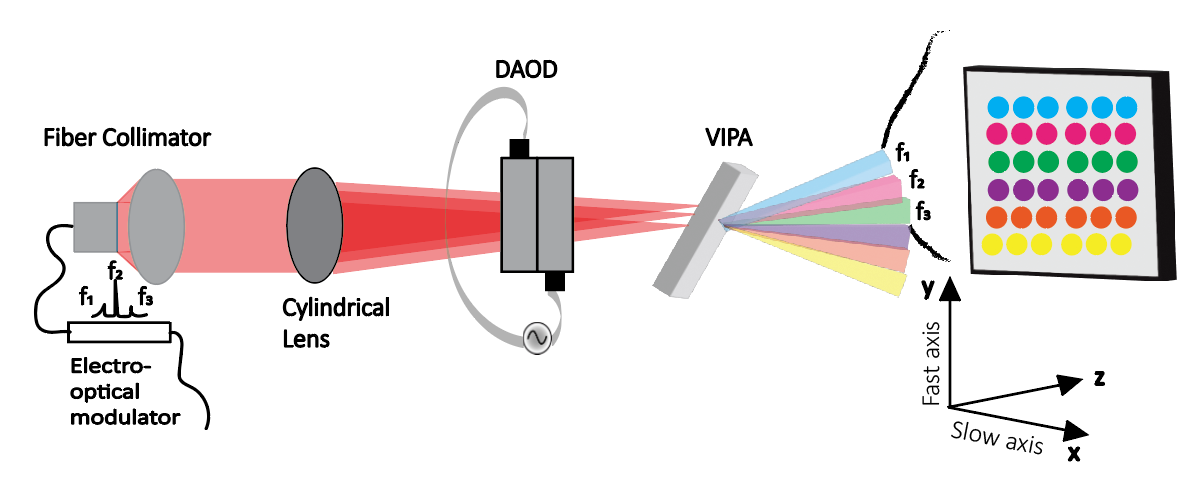}
    \caption{A 2D beam deflector for arbitrary pattern generation, combining a slow-axis pair of AODs (DAOD) and a fast-axis VIPA-EOM system. The DAOD generates the deflection along the $x$-axis (slow axis) and the VIPA separates the optical frequencies generated by the EOM, resulting in a deflection along the $y$-axis (fast axis).}
    \label{fig:1c}
\end{figure*}

Our device achieves its performance by careful design of both the slow and fast axes to achieve the demonstrated speed-up. For the fast axis, we use a Virtually Imaged Phased Array (VIPA) in combination with an electro-optic modulator (EOM), which realizes high-speed control through angular dispersion. For the slow axis, we use a pair of AODs in a counter-propagating configuration that enables the full cancellation of acoustic-lensing effects for linear frequency chirps \cite{wen_novel_2006, kaplan_acousto-optic_2001, friedman_acousto-optic_2000, dan_stamper_kurn_2026}. 
This allows us to reach $\sim 1$ MHz raster refresh rates, much faster than would be possible with a regular AOD at adjustable chirp rate, satisfying the requirement $R_{\text{raster}}>>2\cdot f_{\text{trap}}$.

We first examine the performance of the DAOD and compare it to that of a single AOD to confirm the successful mitigation of the acoustic lensing effect. Our experimental configuration involves a pair of identical AODs (Brimrose TED-150-100-785) with an acoustic velocity $v = 4200$ m/s, a $1.8$ mm aperture, and 100 MHz bandwidth centered around 150 MHz. We access a smaller bandwidth of 36 MHz, limited by the RF block amplifier used. We observe a peak efficiency of 50$\%$ for the AOD and 25$\%$ for the DAOD. 
The access time, $T_a = 2w_0/v$, of the deflected beam to $1/e^2$ of the maximum value was $T_a^{AOD} = (457 \pm 1)$~ns for the AOD and $T_a^{DAOD} = (260 \pm 2)$~ns for the DAOD which is $1.76$ times faster than for the AOD, close to the expected factor of 2 \cite{friedman_acousto-optic_2000}.
The static spatial resolution (ratio of maximum deflection $\Delta x$ to beam waist $w_0$) can be equivalently written as a product of the RF bandwidth and the access time $N_{stat} = \frac{\pi}{4}T_a\Delta F$. We expect $N_{stat}^{AOD} = 33$ for the single AOD and $N_{stat}^{DAOD} = 66$ for the DAOD, given access to the full RF bandwidth $\Delta F$.

\begin{figure}[h]
\includegraphics[width=.95\linewidth]
{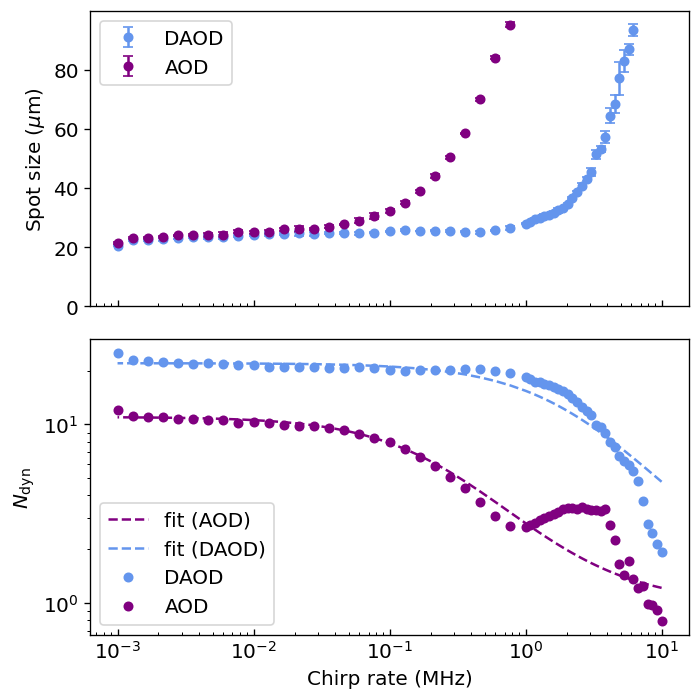}
\caption{(Top) Acoustic defocusing effects for a single AOD and a DAOD undergoing a linear scan across 36 MHz. (Bottom) Dynamic resolution for AOD and DAOD as a function of scan time. Data are fit to theoretical formulae in the text. A resolution of 17 is maintained with the DAOD for linear scans with a sweep rate of up to 1 MHz.}
\label{fig:linear_chirp}
\end{figure}

Next, we evaluate the dynamic resolution while linearly scanning the RF tone and measuring the $1/e^2$ fall time of the optical signal across a knife edge. 
Fig. \ref{fig:linear_chirp} shows the measured beam waist at different scan rates for both the AOD and DAOD setups. We observe acoustic lensing with the AOD as we increase the sweep rate, while the DAOD maintains the original waist size for much faster scans. The AOD data are fit to $N_{dyn}^{AOD} = 1 + N_{stat}/(1+N_{stat} T_a/T_{scan})$, and the DAOD data are fit to $N_{dyn}^{DAOD} = 1+ 2N_{stat}/(1 + T_a/T_{scan})$, where $T_{scan}$ is the scan time, and $T_a$ and $N_{stat}$ are the single AOD parameters. The 3dB roll-off of the dynamic resolution for the AOD occurs at $T_{scan} = 1/( N_{stat} T_a)\sim1/(\Delta F T_a^2)$, limited by both the RF bandwidth and the response time. In contrast, the dynamic resolution of the DAOD doesn't roll-off until $T_{scan} = 1/T_a$. 
Acoustic lensing can also be corrected with a lens, albeit only for a fixed scan rate. The DAOD, however, provides a general solution for any linear scan.

With the current DAOD setup we are able to access a dynamic resolution of $N_{dyn}^{DAOD} =17$ while performing a linear frequency chirp in $1\mu s$ for the "slow" axis. If the DAOD were driven across the full RF band, the dynamic resolution would be $\sim 3$ times larger.

\begin{figure}[h]
\includegraphics[width=.5\textwidth]{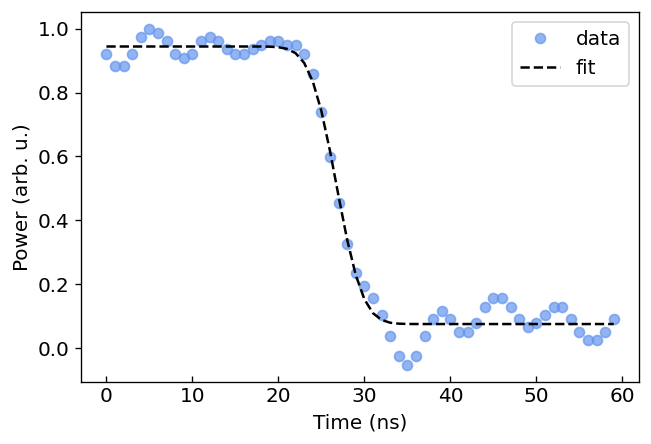}
\caption{VIPA sideband turn-off time measured on a photodetector, $t_{\text{fast}} = (4.8 \pm 0.4$)~ns.}
\label{linear_chirp}
\end{figure}

Next, we test the performance of the combined system by inserting the VIPA into the optical path. We use a $2$mm thick air-spaced VIPA with a 95$\%$ partial reflector coating from Light Machinery (OP-6721-2000-12). We align the VIPA input angle such that we only have a single mode that matches the resonance condition at the output. We turn on a sideband on the fiber EOM (EOSpace 850nm) at 0.1-25.5 GHz.
This results in two visible spots on the camera (Chameleon CM3-U3-13Y3M-CS), one corresponding to the carrier and the other one to the sideband frequency of the EOM. 
We optically relay the VIPA output using a 1:1 4f-setup, with $f = 30$~mm lenses, and place an aperture at the center of the relay such that we filter out the sideband in the image plane. The deflected beam is then focused onto a photo-detector (Newport 1801). We measure an optical switching time of $t_{\text{fast}} = (4.8 \pm 0.4)$~ns on the photo-detector, which is limited by our detector bandwidth (125 MHz) and electric switch. The expected switching time for the VIPA of $t_{\text{fast}} \sim 1$~ns is limited by its linewidth with a FWHM of 1.2 GHz. 

\begin{figure}[h]
\includegraphics[width=.5\textwidth]{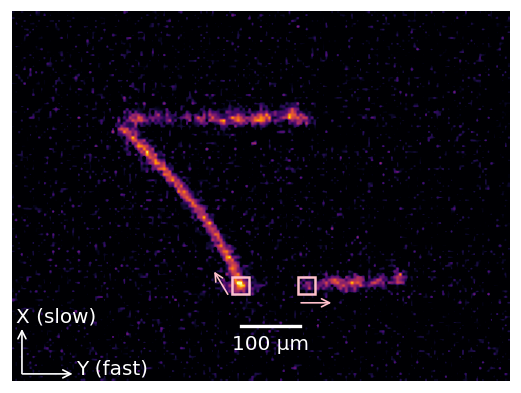}
\caption{An arbitrary optical pattern generated by the raster device capable of moving atoms initially placed in the same row (pink boxes) simultaneously in different directions. This is impossible with a standard 2D AOD device.}
\label{2d_resolution}
\end{figure}

Lastly, we map out the 2D spatial resolution by taking a composite image of many points corresponding to different frequencies within the bandwidth of each device. We were limited by the frequency range of our EOM ($0.1-20$ GHz) and the output power of our RF signal generator, so we could only measure up to 25.5 GHz, and extrapolate the expected resolution for the rest of the frequency range. The expected VIPA resolution is 40 for the full 50 GHz FSR, so we expect to see a resolution of $\sim 20$ by accessing half of it. Due to the beam hitting the imaging lens at slightly different positions we get a $10\%$ reduction in the resolution for the top and bottom rows, which can be improved by using a larger lens. Each individual spot size is measured to be $w_x = (15\pm 1)\mu$m and $w_y = (11.3\pm0.2)\mu$m, extracted with a Lorentzian and a Gaussian fit, respectively.

With full use of the device bandwidth, our combined VIPA+DAOD rasterer has a 2D resolution of approximately 40 x 40 and can raster arbitrary 2D optical potentials with 1 MHz refresh rate, limited by the response time of the DAOD. 
We envision such a device could be used for arbitrary addressing of atoms and atom shuttling in two dimensions. 
The shuttling application imposes stringent requirements: (1) the raster rate must exceed $\sim 1$ MHz to avoid parametric heating \cite{hutzler_2017, savard_heating_1997, zheng_heating_2022}, and (2) atoms must be transported adiabatically in steps of $\sim 100$ nm—the size of an atomic wave packet \cite{bluvstein_transport_2022,evered_high-fidelity_2023}.
Our device meets these requirements, enabling parallel transport of $N$ atoms in arbitrary and independent patterns at a top speed of $\sim 0.1\mu\text{m}/\mu\text{s}$.
Further improving the raster refresh rate would enable even faster transport, as could optimal control techniques. 

With further upgrades, our design could be extended to a device with a 2D resolution of approximately $100 \times 100$. Increasing the resolution of the VIPA would be achieved by adjusting the FSR to make it easier to access the full resolution, and by incorporating a micro-lens array for re-imaging in order to overcome diffractive walk-off which limits the VIPA resolution \cite{jon_simon_2026}. Increasing the resolution of the DAOD can also be attained by substituting the AODs used for one with higher bandwidth, e.g. GPD-800-400-SC from Brimrose. 

The current transmission efficiency of the device, measuring at approximately $\sim 0.02$ (attributed to factors like 0.3 from the fiber EOM, 0.25 for DAOD diffraction efficiency, and 0.25 for the VIPA sideband), is a significant obstacle to implementing this technology. 

Particularly for atom transport,
the low efficiency combined with the EOM power limitations emerges as a primary drawback of the device. 
One could improve the transmission efficiency through the fiber EOM to $60\%$ by heating it up. The sideband could then be interferometrically filtered and a slave laser injection locked to the EOM output. This can then be used to seed a tapered amplifier which could provide Watt-level optical powers. 
The DAOD efficiency will be challenging to improve beyond $25\%$ due to the wide bandwidth of the crystal. However,the VIPA transmission efficiency can be significantly improved by employing I/Q control on the generated sidebands and by focusing more tightly at the input port to reduce any clipping losses. We expect the achievable VIPA transmission efficiency to be $90\%$. Overall, this could result in $\sim 500$mW of optical power at the atoms, which given a $1 \mu$m focus should be enough to generate several hundred movable traps.

In summary, we have introduced a 2D optical rastering device as a versatile tool with exciting implications for various quantum optical applications. Its capabilities extend to facilitating arbitrary individual addressing for Raman transitions \cite{hou_addressing_2024}, enabling precise local qubit rotations, as well as generating arbitrary 2D optical potentials to study the properties of quantum degenerate gases \cite{yan_two-dimensional_2022}. Moreover, the device could significantly accelerate atom rearrangement time scales, and its utility extends to performing geometrically non-local syndrome measurements, a critical requirement for implementing high-rate qLDPC quantum error-correcting codes \cite{preskill_crossing_2023, breuckmann_quantum_2021, bravyi_high-threshold_2024}.

The capabilities of this rastering device extend to numerous other fields requiring high-speed, arbitrary 2D beam steering. In neural imaging, the combination of high refresh rates and arbitrary pattern generation could enable simultaneous volumetric recording from hundreds of neurons with reduced photo-damage compared to raster-scanning approaches \cite{ghosh_2011, yang_2017}. Furthermore, it would allow parallel stimulation of genetically distinct neuronal populations with precise spatiotemporal control for optogenetics applications \cite{boyden_2005, packer_2015}. In LiDAR systems, the purely optical deflection mechanism offers advantages over mechanical scanning methods, enabling compact, solid-state implementations with enhanced reliability and acquisition speeds for autonomous navigation \cite{poulton_2017, amann_2001}. Finally, in free-space optical communication, the rapid beam steering capability could facilitate adaptive beam-forming for atmospheric turbulence compensation and dynamic link establishment in satellite-to-ground or inter-satellite networks \cite{kaushal_2017, khalighi_2014}. The versatility of our rastering approach thus positions it as an enabling technology across diverse photonic applications requiring MHz-rate, arbitrary 2D beam control.

\textit{Note added.} After completion of this work, we became aware of related results reported in Ref. \cite{jon_simon_2026}.

The authors thank Markus Greiner for valuable discussions and Steve Nagle of the T. J. Rodgers Laboratory for providing access to equipment. This material is based upon work supported by the U.S. Department of Energy, Office of Science, National Quantum Information Science Research Centers, Quantum Systems Accelerator (DoE QSA 6945633, grant Research Subcontract No. 7571809). Additional support is acknowledged from DARPA (award $\# 134371-5113608$), the MIT-Harvard Center for Ultracold Atoms (an NSF Frontier Center, award \# PHY-2317134), the NSF Quantum Leap Challenge Institute (award \# 2016244), and the NSF-funded QuSeC-TAQS program (award \# 2326787).

See Supplement 1 for supporting content.

\bibliography{references.bib}

@article{reddy_three-dimensional_2008,
  author  = {Reddy, Gaddum Duemani and Kelleher, Keith and Fink, Rudy and Saggau, Peter},
  title   = {Three-dimensional random access multiphoton microscopy for functional imaging of neuronal activity},
  journal = {Nature Neuroscience},
  year    = {2008},
  volume  = {11},
  number  = {},
  pages   = {713–720},
doi = {https://doi.org/10.1038/nn.2116}
}

@article{wu_multi-mhz_2017,
author = {Jianglai Wu and Anson H. L. Tang and Aaron T. Y. Mok and Wenwei Yan and Godfrey C. F. Chan and Kenneth K. Y. Wong and Kevin K. Tsia},
journal = {Biomed. Opt. Express},
number = {9},
pages = {4160–4171},
publisher = {Optica Publishing Group},
title = {{Multi-MHz} laser-scanning single-cell fluorescence microscopy by spatiotemporally encoded virtual source array},
volume = {8},
month = {Sep},
year = {2017},
url = {https://opg.optica.org/boe/abstract.cfm?URI=boe-8-9-4160},
doi = {10.1364/BOE.8.004160},
}

@article{li_virtually_2021,
author = {Zhi Li and Zihan Zang and H. Y. Fu and Yi Luo and Yanjun Han},
journal = {Appl. Opt.},
number = {8},
pages = {2177–2189},
publisher = {Optica Publishing Group},
title = {Virtually imaged phased-array-based {2D} nonmechanical beam-steering device for {FMCW LiDAR}},
volume = {60},
month = {Mar},
year = {2021},
url = {https://opg.optica.org/ao/abstract.cfm?URI=ao-60-8-2177},
doi = {10.1364/AO.414128},
}

@article{kaplan_acousto-optic_2001,
author = {Ariel Kaplan and Nir Friedman and Nir Davidson},
journal = {Opt. Lett.},
number = {14},
pages = {1078–1080},
publisher = {Optica Publishing Group},
title = {Acousto-optic lens with very fast focus scanning},
volume = {26},
month = {Jul},
year = {2001},
url = {https://opg.optica.org/ol/abstract.cfm?URI=ol-26-14-1078},
doi = {10.1364/OL.26.001078},
}

@article{friedman_acousto-optic_2000,
author = {Nir Friedman and Ariel Kaplan and Nir Davidson},
journal = {Opt. Lett.},
number = {24},
pages = {1762–1764},
publisher = {Optica Publishing Group},
title = {Acousto-optic scanning system with very fast nonlinear scans},
volume = {25},
month = {Dec},
year = {2000},
url = {https://opg.optica.org/ol/abstract.cfm?URI=ol-25-24-1762},
doi = {10.1364/OL.25.001762},
}

@misc{contreras_neutral_2019,
  author       = {Contreras, A. K. and Bernien, Hannes and Schwartz, S. and Levine, H. J. and Omran, A. and
  Lukin, Mikhail D. and Vuletic, Vladan and Endres, Manuel and Greiner, Markus and Pichler, Hannes and
  Zhou, L. and Wang, S. and Choi, S. and Kim, D. and
  Zibrov, A. S.},
  title        = {{Neutral Atom Quantum Information Processor}},
  howpublished = {Canadian patent CA3102913A1 (published application)},
  number       = {CA3102913A1},
  year         = {2018}, 
  url          = {https://patents.google.com/patent/CA3102913A1/en},
}

@article{yan_two-dimensional_2022,
  title = {{Two-Dimensional Programmable Tweezer Arrays of Fermions}},
  author = {Yan, Zoe Z. and Spar, Benjamin M. and Prichard, Max L. and Chi, Sungjae and Wei, Hao-Tian and Ibarra-Garc\'{\i}a-Padilla, Eduardo and Hazzard, Kaden R. A. and Bakr, Waseem S.},
  journal = {Phys. Rev. Lett.},
  volume = {129},
  issue = {12},
  pages = {123201},
  numpages = {6},
  year = {2022},
  month = {Sep},
  publisher = {American Physical Society},
  doi = {10.1103/PhysRevLett.129.123201},
  url = {https://link.aps.org/doi/10.1103/PhysRevLett.129.123201}
}

@misc{preskill_crossing_2023,
  author       = {Preskill, John.},  
  title        = {{Crossing the Quantum Chasm from NISQ to Fault Tolerance}},  
  howpublished = {{Keynote address at Q2B 2023 Conference}},  
  year         = {2023},  
  url          = {https://quantumfrontiers.com/2023/12/09/crossing-the-quantum-chasm-from-nisq-to-fault-tolerance/},  
}

@article{breuckmann_quantum_2021,
  title = {{Quantum Low-Density Parity-Check Codes}},
  author = {Breuckmann, Nikolas P. and Eberhardt, Jens Niklas},
  journal = {PRX Quantum},
  volume = {2},
  issue = {4},
  pages = {040101},
  numpages = {19},
  year = {2021},
  month = {Oct},
  publisher = {American Physical Society},
  doi = {10.1103/PRXQuantum.2.040101},
  url = {https://link.aps.org/doi/10.1103/PRXQuantum.2.040101}
}

@article{hutzler_2017,
  doi = {10.1088/1367-2630/aa5a3b},
  title = {Eliminating light shifts for single atom trapping},
  author = {Hutzler, Nicholas R. and Liu, Lee R. and Yu, Yichao and Ni, Kang-Kuen},
  journal = {{New J. Phys.}},
  volume = {19},
  pages = {023007},
  year = {2017}
}

@article{savard_heating_1997,
issn = {1050-2947},
author = {Savard, T. A. and Savard, T A and O’Hara, K M and Thomas, J E},
address = {New York, N.Y. :},
journal = {{Phys. Rev. A}},
lccn = {90656533},
number = {2},
pages = {R1095-R1098},
publisher = {Published by the American Physical Society through the American Institute of Physics,},
title = {Laser-noise-induced heating in far-off resonance optical traps},
volume = {56},
year = {1997-8-1},
doi = {10.1103/PhysRevA.56.R1095}
}

@article{zheng_heating_2022,
author = {Zheng, Ningxuan and Liu, Wenliang and Wu, Jizhou and Li, Yuqing and Sovkov, Vladimir and Ma, Jie},
title = {Parametric Excitation of Ultracold Sodium Atoms in an Optical Dipole Trap},
journal = {Photonics},
volume = {9},
year = {2022},
number = {7},
pages = {442},
doi = {10.3390/photonics9070442}
}

@misc{jon_simon_2026,
      title={A 10 Megahertz Spatial Light Modulator}, 
      author={Xin Wei and Zeyang Li and Abhishek V. Karve and Adam L. Shaw and David I. Schuster and Jonathan Simon},
      year={2026},
      eprint={2601.08906},
      archivePrefix={arXiv},
      primaryClass={quant-ph},
      url={https://arxiv.org/abs/2601.08906}, 
}

@misc{dan_stamper_kurn_2026,
      title={Astigmatism-free 3D Optical Tweezer Control for Rapid Atom Rearrangement}, 
      author={Yue-Hui Lu and Nathan Song and Tai Xiang and Jacquelyn Ho and Tsai-Chen Lee and Zhenjie Yan and Dan M. Stamper-Kurn},
      year={2026},
      eprint={2510.11451},
      archivePrefix={arXiv},
      primaryClass={physics.optics},
      url={https://arxiv.org/abs/2510.11451}, 
}

@article{wu_kilohertz_2020,
  author  = {Wu, J. and Liang, Y. and Chen, S.},
  title   = {Kilohertz two-photon fluorescence microscopy imaging of neural activity in vivo},
  journal = {Nature Methods},
  year    = {2020},
  volume  = {17},
  pages   = {287--290},
  doi     = {10.1038/s41592-020-0762-7}
}

@article{wu_ultrafast_2017,
  author  = {Wu, J. L. and Xu, Y. Q. and Xu, J. J.},
  title   = {Ultrafast laser-scanning time-stretch imaging at visible wavelengths},
  journal = {Light Sci Appl},
  year    = {2017},
  volume  = {6},
  pages   = {e16196},
  doi     = {10.1038/lsa.2016.196}
}

@article{lai_high-speed_2021,
  author  = {Lai, Q. T. K. and Yip, G. G. K. and Wu, J.},
  title   = {High-speed laser-scanning biological microscopy using {FACED}},
  journal = {Nat Protoc},
  year    = {2021},
  volume  = {16},
  pages   = {4227--4264},
  doi     = {10.1038/s41596-021-00576-4}
}

@article{bluvstein_logical_2024,
  author  = {Bluvstein, D. and Evered, S. J. and Geim, A. A.},
  title   = {Logical quantum processor based on reconfigurable atom arrays},
  journal = {Nature},
  year    = {2024},
  volume  = {626},
  pages   = {58--65},
  doi     = {10.1038/s41586-023-06927-3}
}

@article{evered_high-fidelity_2023,
  author  = {Evered, S. J. and Bluvstein, D. and Kalinowski, M.},
  title   = {High-fidelity parallel entangling gates on a neutral-atom quantum computer},
  journal = {Nature},
  year    = {2023},
  volume  = {622},
  pages   = {268--272},
  doi     = {10.1038/s41586-023-06481-y}
}

@article{manetsch_tweezer_2024,
  author  = {Manetsch, H. J. and Nomura, G. and Bataille, E.},
  title   = {A tweezer array with 6,100 highly coherent atomic qubits},
  journal = {Nature},
  year    = {2025},
  volume  = {647},
  pages   = {60--67},
  doi     = {10.1038/s41586-025-09641-4}
}

@article{wen_novel_2006,
  author  = {Wen, T. and Zhuang, Z. and Wei, J.},
  title   = {A novel method to improve spatial resolution of acousto-optic deflector},
  journal = {Optoelectr. Lett.},
  year    = {2006},
  volume  = {2},
  pages   = {34--36},
  doi     = {10.1007/BF03033588}
}

@article{bravyi_high-threshold_2024,
  author  = {Bravyi, S. and Cross, A. W. and Gambetta, J. M.},
  title   = {High-threshold and low-overhead fault-tolerant quantum memory},
  journal = {Nature},
  year    = {2024},
  volume  = {627},
  pages   = {778--782},
  doi     = {10.1038/s41586-024-07107-7}
}

@article{bluvstein_transport_2022,
  author  = {Bluvstein, D. and Levine, H. and Semeghini, G.},
  title   = {A quantum processor based on coherent transport of entangled atom arrays},
  journal = {Nature},
  year    = {2022},
  volume  = {604},
  pages   = {451--456},
  doi     = {10.1038/s41586-022-04592-6}
}

@article{hou_addressing_2024,
  author  = {Hou, Y. H. and Yi, Y. J. and Wu, Y. K.},
  title   = {Individually addressed entangling gates in a two-dimensional ion crystal},
  journal = {Nat Commun},
  year    = {2024},
  volume  = {15},
  pages   = {9710},
  doi     = {10.1038/s41467-024-53405-z}
}

@article{ghosh_2011,
  author  = {Ghosh, K. and Burns, L. and Cocker, E.},
  title   = {Miniaturized integration of a fluorescence microscope},
  journal = {Nat. Methods},
  year    = {2011},
  volume  = {8},
  pages   = {871--878},
  doi     = {10.1038/nmeth.1694}
}

@article{yang_2017,
  author  = {Yang, W. and Yuste, R.},
  title   = {In vivo imaging of neural activity},
  journal = {Nat. Methods},
  year    = {2017},
  volume  = {14},
  pages   = {349--359}
}

@article{boyden_2005,
  author  = {Boyden, E. S. and Zhang, F. and Bamberg, E. and Nagel, G. and Deisseroth, K.},
  title   = {Millisecond-timescale, genetically targeted optical control of neural activity},
  journal = {Nat. Neuroscience},
  year    = {2005},
  volume  = {8},
  pages   = {1263--1268}
}

@article{packer_2015,
  author  = {Packer, A. M. and Russell, L. E. and Dalgleish, H. W. P. and H{\"a}usser, M.},
  title   = {Simultaneous all-optical manipulation and recording of neural circuit activity with cellular resolution in vivo},
  journal = {Nat. Methods},
  year    = {2015},
  volume  = {12},
  pages   = {140--146}
}

@article{amann_2001,
  author  = {Amann, M. C. and Bosch, T. and Lescure, M. and Myllyla, R. and Rioux, M.},
  title   = {Laser ranging: a critical review of usual techniques for distance measurement},
  journal = {Optical Engineering},
  year    = {2001},
  volume  = {40},
  pages   = {10--19}
}

@article{poulton_2017,
  author  = {Poulton, C. V. and Yaacobi, A. and Cole, D. B. and Byrd, M. J. and Raval, M. and Vermeulen, D. and Watts, M. R.},
  title   = {Coherent solid-state {LIDAR} with silicon photonic optical phased arrays},
  journal = {Optics Letters},
  year    = {2017},
  volume  = {42},
  pages   = {4091--4094}
}

@article{kaushal_2017,
  author  = {Kaushal, H. and Kaddoum, G.},
  title   = {Optical communication in space: Challenges and mitigation techniques},
  journal = {IEEE Communications Surveys \& Tutorials},
  year    = {2017},
  volume  = {19},
  pages   = {57--96}
}

@article{khalighi_2014,
  author  = {Khalighi, M. A. and Uysal, M.},
  title   = {Survey on free space optical communication: A communication theory perspective},
  journal = {IEEE Communications Surveys \& Tutorials},
  year    = {2014},
  volume  = {16},
  pages   = {2231--2258}
}

\setcounter{figure}{0} 
\renewcommand{\thefigure}{S\arabic{figure}} 
\renewcommand{\theequation}{S\arabic{equation}}

\section*{Supplementary Materials}

\section*{S1. Experimental setup}
\label{sec:appdxA}
Fig. \ref{linear_chirp} shows a detailed schematic of the entire optical setup used to calibrate and characterize the device.

\begin{figure}[h]
\includegraphics[angle = 0, width=\linewidth, trim = {0cm 0cm 0cm 0cm}]{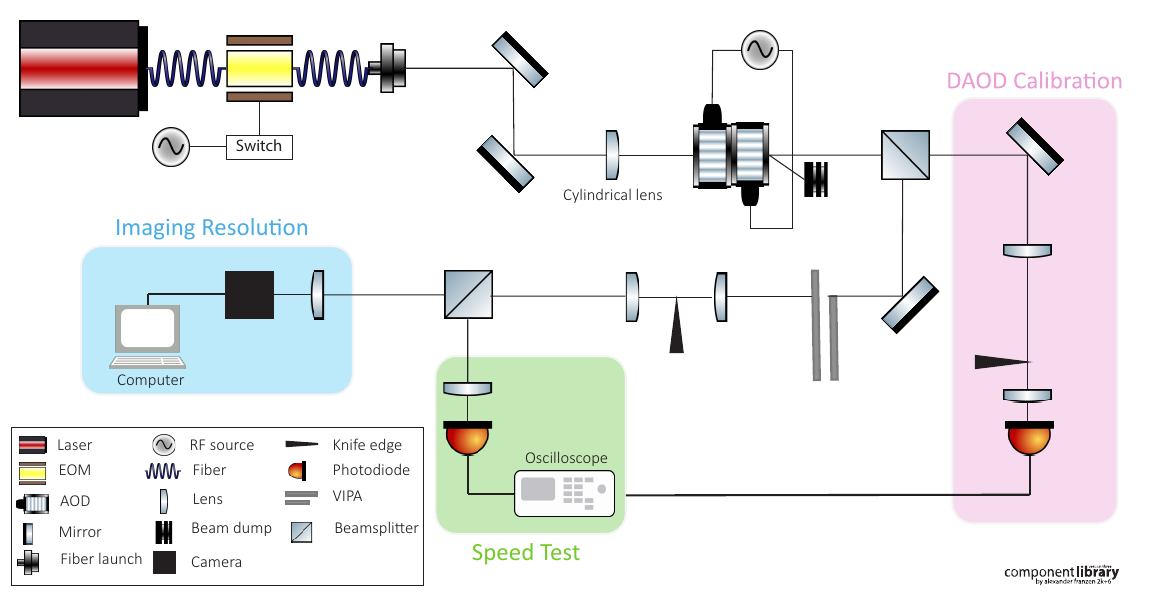}
\caption{Experimental setup of 2D rastering device, including paths for calibrating the DAOD system, determining the resolution, and testing the switching speed of VIPA sidebands.}
\label{linear_chirp}
\end{figure}

\section*{S2. Design considerations}
\label{sec:appdxB}
\begin{figure}[h]
    \centering
    \includegraphics[width=0.5\linewidth, trim = {3.5cm 0 4.5cm 0}]{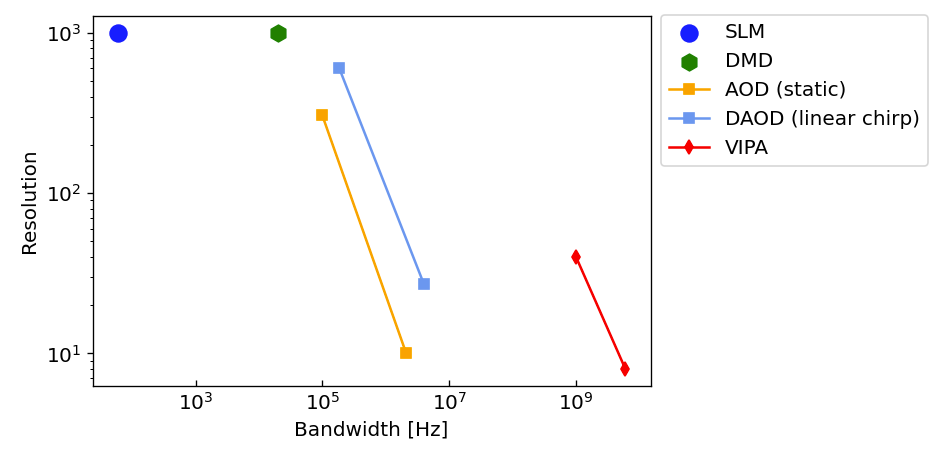}
    \caption{Comparison of various optical deflectors used to evaluate the resolution-bandwidth tradeoff for the slow and fast axes of the rastering device. The combination of DAOD for the slow axis and VIPA for the fast axis fulfills the speed and resolution requirements.}
    \label{fig:fig6}
\end{figure}

In Fig. \ref{fig:fig6}, we survey available 1D optical deflectors and compare their bandwidth and resolution to determine the best candidates for the "fast" and "slow" axes. 

To understand the speed requirements, consider the case of addressing an arbitrary number of atoms, $N$, on a 2D grid of $n\times n$ sites. With a classical crossed AOD system, addressing all atoms requires up to $N\cdot T_a$ time, since the AOD must sequentially switch each tone on and off \cite{contreras_neutral_2019}. Replacing each axis with a DAOD yields only a two-fold improvement with an effective access time of $T_a/2$, because the counter-propagating RF waves only need to travel half the beam diameter. We cannot take advantage of any further dynamic speed-up here, because if one axis does a linear scan, the other axis controlled by a second DAOD cannot keep up unless the number of atoms is extremely low ($\sim 1$). An AOD-VIPA configuration also faces limitations: even with the fast VIPA, the AOD must scan slowly enough to maintain a dynamic resolution of $N_{dyn}=n$ along the horizontal axis, spending at least $T_a$ per column and thus requiring $N\cdot T_a$ total time to address $N$ atoms (aside from rare cases where all the atoms are in a line sharing the same coordinate along the slow axis). Our DAOD-VIPA configuration overcomes this bottleneck. Because the acoustic lensing effect is completely canceled for linear chirps in the DAOD, we can address all $n$ columns in time $t\sim T_a$, limited only by the DAOD's access time rather than by $N\cdot T_a$. This approach additionally requires the VIPA to switch rapidly enough to keep pace with the slow axis, imposing a switching time requirement of $t_{fast}\leq T_a/n$. This condition is readily achievable for $n\leq 500$, enabling our device to provide an $\sim N$-fold speed advantage over existing deflector systems for addressing applications.

As in the addressing case, crossed AOD or DAOD systems require a time up to $t_{\text{shuttle}}\sim \mathcal{O}(N\cdot T_a)$ to shuttle $N$ atoms in arbitrary directions, since each atom must be moved sequentially. An AOD-VIPA configuration faces a similar limitation: the ``slow" axis must move slowly enough to avoid smearing out the tweezer, also requiring $t_{\text{shuttle}}\sim\mathcal{O}(N\cdot T_a)$.
The DAOD-VIPA system, in contrast, can address an arbitrary pattern of $N$ atoms during each rastering cycle ($\sim 1~\mu\text{s}$), eliminating the $N$-dependence and reducing the required time to $t_{\text{shuttle}}\sim\mathcal{O}(T_a)$ for the entire motion. Note that this analysis focuses on the addressing overhead; the actual transport time per atom is identical across all systems. By enabling parallel addressing of $N$ atoms per raster cycle, our device achieves an $N$-fold speed improvement for shuttling operations. The raster refresh rate sets the speed of atom motion, $v\sim 0.1\mu m\cdot R_{\text{raster}}$, so an improvement in the device's speed would result in faster motion.

\section*{S3. Counter-propagating configuration of AOD}
\label{sec:appdxC}
An AOD deflects a beam by an angle, $\theta = \lambda \frac{f}{v}$, directly proportional to the applied RF frequency $f$ and inversely proportional to the acoustic speed in the crystal $v$ \cite{friedman_acousto-optic_2000}. We will refer to two key properties of AODs: (i) the access time, $T_a = \frac{2w_0}{v}$ ($w_0$ is the input beam waist), and (ii) the resolution, $N = \frac{\Delta\theta}{\delta \theta}$ ($\delta\theta$ is the spot size, and $\Delta\theta$ is the full deflection range).

The access time imposes a constraint in scanning applications of AODs introducing an acoustic lensing effect. Consider a linear chirp over the full RF bandwidth in $T_{scan}\sim 10 \cdot T_a$. Then within any given infinitesimal time $dt$ there is a range of driving frequencies $dF$ present inside the crystal. This results in the output beam being deflected in a range of angles $d\theta$, which effectively lowers the number of resolvable spots. The angular spread of the deflected beam for the linear chirp case is then:
\begin{equation}
    \delta \theta_{chirp} = \frac{2}{\pi} \frac{\lambda}{w_0} + \frac{\lambda}{v}dF,
\end{equation}
which we use to define the dynamic resolution $N_{dyn} = \frac{\Delta\theta}{\delta\theta_{chirp}}$. Using the relation $dF = \frac{\Delta F}{T_{scan}}dt$, it can be rewritten as:

\begin{equation}
    N_{dyn}^{AOD} = \frac{N}{1+N\frac{T_a}{T_{scan}}}+1.
\end{equation}

where $N$ and $T_a$ are the single AOD static resolution and access time, respectively. The defocusing due to the AOD lensing effect is equivalent to a cylindrical lens of focal length, $f_{AOD} = \frac{a^2 v^2}{\lambda \alpha}$, where $a$ is a constant dependent on the beam profile ($\sim 1.34$ for $TEM_{00}$) and $\alpha = \Delta F/ T_{scan}$ is the frequency chirp rate. The AOD focal length $f_{AOD}$ is usually large ($\sim$ m) compared to the focal length of the lenses we use in the setup $f_{obj}$ ($\sim$ mm), so we can approximate the focal shift as $f_{shift} \approx -\frac{f_{obj}^2}{f_{AOD}}$.

If one adds a second AOD with a counter-propagating acoustic wave, this spot expansion is suppressed and the resolution is almost quadrupled \cite{wen_novel_2006, friedman_acousto-optic_2000}. Furthermore, the 3dB roll-off of the dynamic resolution happens at much shorter scan times, $T_{scan}\sim T_a$, resulting in:
\begin{equation}
    N_{dyn}^{DAOD} = \frac{2N}{1+\frac{T_a}{T_{scan}}} + 1.
\end{equation}

The proposed counter-propagating double AOD (DAOD) setup relies on the cancellation of the acoustic lensing effect of a single AOD for linear scans, thus improving the dynamic resolution. The input beam is deflected up by some angle $\theta_B$ after the first AOD and receives a momentum kick parallel to the acoustic-wave propagation resulting in an optical frequency up-shift ($\frac{\omega}{2\pi}+F_0$) equivalent to the frequency of the RF tone ($F_0$) used to drive the AOD. Then, the +1 diffracted order is transmitted through a second counter-propagating AOD where it is deflected up again by $\theta_B$ and receives a momentum kick anti-parallel to the acoustic wave propagation resulting in a frequency down-shift. In the absence of a scan, the static resolution of DAODs is $N_{DAOD}^{stat} = \frac{2\Delta \theta}{\delta \theta} = 2N_{AOD}^{stat}$.

\end{document}